\documentstyle[aps,amsfonts,prbbib]{revtex}

\newcommand{\Ri}{{\bf R}}
\newcommand{\seR}{{\Bbb R}}
\newcommand{\seA}{{\Bbb A}}
\newcommand{\wek}[2]{\newcommand{#1}{\mbox{\bf #2}}}
\wek{\vR}{R} \wek{\vrr}{r} \wek{\vA}{A} \wek{\vH}{H} \wek{\vp}{p}
\newcommand{\ii}{\mbox{\rm i}}
\newcommand{\sig}[1]{\sum_{#1=1}^{3}\,}

\begin{document}
 
 \title{Local gauge and magnetic translation groups}
 \author{Wojciech Florek\cite{AAA}}
 \address{A.~Mickiewicz University, Institute of Physics, ul.~Umultowska 
85, 61--614 Pozna\'n, Poland}
 \date{1 August, 1997}

\maketitle

\begin{abstract}

 The magnetic translation group was introduced as a set of operators 
 $T(\vR)=\exp[-\ii\vR\cdot (\vp-e\vA/c)/\hbar].$ However,
these operators commute with the Hamiltonian for an electron in a periodic
potential and a uniform magnetic field if the vector potential \vA\ (the
gauge) is chosen in a symmetric way. It is showed that a local gauge field 
 $\vA_{\Ri}(\vrr)$ on a crystal lattice leads to
operators, which commute with the Hamiltonian for any (global) gauge field
$\vA=\vA(\vrr)$. Such choice of the local gauge determines a
factor system $\omega(\vR,\vR')= T(\vR)T(\vR')
 T(\vR+\vR')^{-1}$, which depends on a global gauge only. Moreover, 
for any potential \vA\ a commutator $T(\vR)T(\vR')T(\vR)^{-1}
T(\vR')^{-1}$ depends only on the magnetic field and not on the gauge.

 \end{abstract}


\section{Introduction} 
 The behavior of electrons in crystalline (periodic) potentials in the
presence of a constant (external) magnetic field has been studied since the
thirties in many papers. \cite{landau,peierls}) In the sixties
Brown \cite{brown} and Zak \cite{zak1,zak2} independently introduced and
investigated the so-called {\em magnetic translation groups}.  Their results
have been lately applied to a~problem of the quantum Hall
effect~\cite{dana1,dana2} and relations with the Weyl--Heisenberg group have
been also studied. \cite{zak} Some interesting results have been presented
lately by Geyler and Popov. \cite{geypop}

The Hamiltonian for an electron in a periodic potential $V(\vrr)$ and a
uniform magnetic field (described by the vector potential \vA) is given
as \cite{brown,zak1}
 \begin{equation}\label{hamil}
  {\cal H}\,=\,\pi^2/2m+V(\vrr)\,,\qquad\mbox{where}\qquad   
  \pi\,=\,\vp+e\vA/c 
 \end{equation}
 is the (vector) operator of the kinetic momentum. Brown introduced a
projective representation of the translation group in the following form
\begin{equation}\label{brownd}
 T(\vR)\,=\,\exp[-\ii(\vp-e\vA/c)\cdot\vR/\hbar]\,.
\end{equation}
 These operators commute with the Hamiltonian (\ref{hamil}) if the vector
potential~\vA\ fulfills the following condition \cite{brown,zak1}
 \begin{equation} 
  \partial A_j/\partial x_k+\partial A_k/\partial x_j\,=\,0;\qquad 
   {\rm for\ }  j,k\,=\,1,2,3\,.
\end{equation}
 This relation holds, for example, for the gauge $\vA(\vrr)=
(\vH\times\vrr)/2$, which was used by Brown and Zak. \cite{brown,zak1} 
On the other hand, this condition is not satisfied by the Landau gauge
$\vA(\vrr)=[-x_2H_3,0,0]$ (for $\vH=[0,0,H_3]$), which is used in
many papers. \cite{WZ} 

The aim of this paper is to find such a gauge $\vA'$ that: (i)
$\nabla\times \vA'=-\vH$; (ii) operators $T'(\vR)=
 \exp[-\ii(\vp+\vA'e/c)\cdot\vR/\hbar]$ commute with the
Hamiltonian (\ref{hamil}); (iii) a factor system $\omega(\vR,\vR')=
T(\vR)T(\vR') T(\vR+\vR')^{-1}$ depends only on a global
gauge \vA, which defines the magnetic field (and the generalized momentum
$\pi$ in (\ref{hamil})). It should be underline that only the constant
magnetic field \vH\ is considered. It occurs that these conditions are
satisfied by a {\em local}\/ gauge, i.e.\ an actual form of
 $\vA'(\vrr)$ depends on a lattice vector \vR.

\section{Solution}
 For the constant magnetic field $\vH=[H_1,H_2,H_3]$ the vector
potential (gauge) $\vA=[A_1,A_2,A_3]$ can be chosen as a linear function
of $\vrr=[x_1,x_2,x_3]$ and can be written as
 \begin{equation}
 A_j\,=\,\sig{k}a_{jk}\,x_k\,,\qquad 
 \mbox{with} \qquad a_{jk}\in\seR\,,
 \quad a_{jj}\,=\,0\,.
 \end{equation}
 Introducing a matrix $\seA=(a_{jk})$ it can be written as $\vA=\seA\vrr$.
Therefore, the magnetic field \vH\ is expressed by the matrix elements
$a_{jk}$ as follows 
 \begin{equation}
 H_j=-\sig{k,l}\varepsilon_{jkl}\, a_{kl}\,,
 \end{equation}
 what means that \vH\ is related to antisymmetrized matrix $\seA$.

The definition (\ref{brownd}) of operators $T(\vR)$ can be rewritten as
\begin{equation}\label{browndx}
 T(\vR)\,=\,\exp(-\ii\pi'\cdot\vR/\hbar)\,,
\end{equation}
 where
 $$
 \pi'\,=\,\vp+e\vA'/c\qquad\mbox{and} \qquad \vA'(\vrr)\,=\,-\vA(\vrr)\,,
 $$
 so $\nabla\times\vA'=-\nabla\times\vA=-\vH$. Let us consider
$\vA'=\vA-\vH\times\vrr=\seA^T\vrr$, i.e.
 \begin{equation}\label{afirst}
 A'_j\,=\,A_j-\sig{k,l}\varepsilon_{jkl}\,H_k x_l\,=\,\sig{k}a_{kj}x_k\,.
 \end{equation}
 It is easy to note that $\nabla\times\vA'=-\vH$. For example, assuming the
Landau gauge (for $\vH=[0,0,H_3]$) to be given as $\vA=[-x_2H_3,0,0]$ one
obtains $\vA'=[0,-x_1H_3,0]$, whereas the symmetric gauge $(\vH\times\vrr)/2$ 
yields $\vA'=-(\vH\times\vrr)/2$. We have to check whether the operators
$T(\vR)$ determined by the gauge (\ref{afirst}) commute with the Hamiltonian.
It suffices to calculate commutators $[\pi_j,\pi'_k]$ for $j,k=1,2,3$, for
which one obtains
 $$
 [\pi_j,\pi'_k]\,=\,[-\ii\hbar\partial_j+eA_j/c,-\ii\hbar\partial_k+eA'_k/c]
 \,=\,-\ii e\hbar([\partial_j,A'_k]+[A_j,\partial_k])/c\,=\,0\,.
 $$

 To find a factor system of the above determined (projective) representations
one has to calculate commutators $[X_j\pi'_j,X'_k\pi'_k]$
($\vR=[X_1,X_2,X_3]$): 
 \begin{eqnarray*}
 [X_j\pi_j,X'_k\pi'_k]&=&
  X_jX'_k[-\ii\hbar\partial_j+eA'_j/c,-\ii\hbar\partial_k+eA'_k/c]\\
 &=&-\ii X_jX'_k e\hbar/c(\partial_j\,A'_k -\partial_k\,A'_j)
 \;=\;-\ii X_jX'_k e\hbar/c(a_{jk}-a_{kj})\,.
 \end{eqnarray*}
 On the other hand we have
 $$
 (\vR\times\vR')\cdot\vH\,=\,-\sig{l}\left(
 \sig{j,k}\varepsilon_{ljk}\,X_jX'_k\right)
 \left(\sig{p,q}\varepsilon_{lpq}\, a_{pq}\right)\,=\,
 -\sig{j,k}\,X_jX'_k\, (a_{jk}-a_{kj})
 $$
 and, therefore,
 $$
 T(\vR)T(\vR')\,=\,
  T(\vR+\vR')\exp[-\ii(e/\hbar c)(\vR\times\vR')\cdot\vH/2]\,.
 $$
 It is interesting that this result does not depend on the chosen gauge \vA\
(and $\vA'$).

 Let us consider now a {\em local}\/ gauge determined as
$\vA'_{\Ri}(\vrr)=\seA\,(\vrr+\vR/2)$, i.e.
 \begin{equation}\label{asecond}
 (A'_{\Ri})_j(\vrr)\,=\,\sig{k}a_{kj}(x_k+X_k/2)\,.
  \end{equation}
  Similar, but a bit more tedious, calculations lead to the following
results:\\
\hspace*{\parindent} 1. $\nabla\times\vA'_{\Ri}=-\vH$;\\
\hspace*{\parindent} 2. Operators $T(\vR)$ determined by this gauge commute
with the Hamiltonian (\ref{hamil});\\
\hspace*{\parindent} 3. The projective representation $T(\vR)$ is
characterized by a factor system 
 \begin{equation}\label{factor}
 \omega(\vR,\vR')\,=\,T(\vR)T(\vR')T(\vR+\vR')^{-1}\,=\,
 \exp[(-\ii e/\hbar c)(\vR\cdot\vA(\vR'))]\,. 
 \end{equation}

Note that a scalar product in the last equation can be also written as
a bilinear form 
 \begin{equation}\label{scalar}
 \vR\cdot\vA(\vR')\,=\,\sig{j,k}a_{jk} X_j X'_k\,=\,\vR\cdot\seA\,\vR'
 \,=\,\seA^T\,\vR\cdot\vR'\,,
 \end{equation}
 i.e.\ it is fully determined by the matrix $\seA$. It should be stressed
that calculating $\omega(\vR,\vR')$ one has to take into account that
 $$
 T(\vR+\vR')\,=\,\exp[-\ii(\vp+e\vA'_{\Ri+\Ri'}/c)\cdot(\vR+\vR')/\hbar]\,.
 $$
 The obtained factors (\ref{factor}) allow to find the commutator 
 $T(\vR)T(\vR')T(\vR)^{-1}T(\vR')^{-1}$ as
 $$
 \omega(\vR,\vR')\omega(\vR',\vR)^{-1}\,=\,
 \exp[(-\ii e/\hbar c)\vR\cdot(\seA-\seA^T)\vR')]\,.
 $$
 Applying (\ref{scalar}) this result can be also written as
 $$
 \exp\left[(-\ii e/\hbar c)\left(
 \sig{j,k} X_j X'_k (a_{jk}-a_{kj}) \right)\right]\,.
 $$
 Hence, we have showed that 
 $$
 T(\vR)T(\vR')T(\vR)^{-1}T(\vR')^{-1}\,=\,
 \exp[(\ii e/\hbar c)\Phi]\,,
 $$
 where $\Phi=(\vR\times\vR')\cdot\vH$ is a magnetic flux through the cell
spanned by lattice vectors \vR\ and $\vR'$. 

\section{Conclusion}
  It was showed that a projective representation (\ref{browndx}) of the
translation group determined by the {\em local}\/ gauge (\ref{asecond})
has the following properties:\\
\hspace*{\parindent} 1. Operators $T(\vR)$ commute with the Hamiltonian
(\ref{hamil}).\\
\hspace*{\parindent} 2. The factor system $\omega(\vR,\vR')$ depends on the
{\em global}\/ gauge \vA, i.e.\ on the matrix $\seA$.\\
\hspace*{\parindent} 3. The commutator of (magnetic) translations $T(\vR)$
and $T(\vR')$ depends only on the magnetic field \vH.

%aaaaaaaaaaaaaaa

\end{document}